\documentclass[conference]{IEEEtran}
\IEEEoverridecommandlockouts
\usepackage{cite}
\usepackage{amsmath,amssymb,amsfonts}
\usepackage{algorithmic}
\usepackage{graphicx}
\usepackage{textcomp}
\usepackage{xcolor}
\usepackage[T1]{fontenc} 
\def\BibTeX{{\rm B\kern-.05em{\sc i\kern-.025em b}\kern-.08em
    T\kern-.1667em\lower.7ex\hbox{E}\kern-.125emX}}
\usepackage{fancyhdr}

\begin{document}

\title{Best implementations of quaternary adders }

\author{\IEEEauthorblockN{ Daniel Etiemble}
\IEEEauthorblockA{\textit{Computer Science Laboratory (LRI)} \\
\textit{Paris Saclay University}\\
Orsay, France \\
de@lri.fr}

}
\maketitle

\begin{abstract}
The implementation of a quaternary 1-digit adder composed of a 2-bit binary adder, quaternary to binary decoders and binary to quaternary encoders is compared with several recent implementations of quaternary adders. This simple implementation outperforms all other implementations using only one power supply. It is equivalent to the best other implementation using three power supplies. The best quaternary adder using a 2-bit binary adder, the interface circuits between quaternary and binary levels are just overhead compared to the binary adder. This result shows that the quaternary approach for adders use more transistors, more chip area and more power dissipation than the corresponding binary ones.

\end{abstract}


\section{Introduction}\label{sec1}
Many designs of quaternary adders have been proposed in the recent years. Most of these papers are based on simulations using parameters of CNTFET technology. The recent most significant ones are \cite{Ebrahimi}\cite{Moaiyeri} \cite{Roosta}.:
\begin{itemize}  
\item \cite{Ebrahimi} only uses one power supply.
\item The quaternary half adder presented in \cite{Moaiyeri} uses 3 power supplies, even if the technique used to get the intermediate power supplies is not precised.
\item \cite{Roosta} presents both single-supply and 3 supplies versions.
\end{itemize}
In this paper, we propose a new design of quaternary adders using the same assumptions as in these three papers. This design leads to the most efficient implementation in term of transistor count. 

\section{Methodology}
\subsection{Why CNTFET technology?}
This technology uses field-effect transistors that use a single carbon nanotube or an array of carbon nanotubes as the channel material instead of bulk silicon in the traditional MOSFETs. The MOSFET-like CNTFETs having p and n types look the most promising ones. The technology has advantages and drawbacks:
\begin{itemize}
\item CNTFETs have variable threshold voltages (according to the inverse function of the diameter). This is a big advantage compared to CMOS for which different masks are needed to get different threshold voltages. 
\item Among advantages, high electron mobility, high current density, high tranductance can be quoted.
\item Lifetime issues, reliability issues, difficulties in mass production and production costs are quoted as disadvantages.
\item CNTFET technology is far from being a mature one. In 2019, a 16-bit RISC microprocessor has been built with 14,000 CNFET transistors \cite{Hills}. While this is an advance for CNTFET technology, we may observe that the Intel 8086 CPU, which was a 16-bit microprocessor, has been launched in 1978 with 29,000 transistors, more than 40 years ago!
\end{itemize} 
However, as CMOS circuits and CNTFET ones have basically the same circuit styles, CNTFETs can be used to propose a new implementation of quaternary adders and compare it with previous published proposals.

\subsection {Comparing different implementations of quaternary adders}
The transistor count is used to compare different implementations of quaternary adders. As comparisons are done by using the same technology and the same operators, the transistor count  is significant as it is very doubtful that more transistors could lead to: 
\begin{itemize}
\item	less interconnects
\item	reduced chip area
\item	reduced power dissipation
\item	reduced propagation delays
\item	Etc.
\end{itemize}

\section{Quaternary circuits}
\subsection{Four different levels}
While binary circuits have 0 and 1 levels, quaternary circuits have four levels 0 < 1 < 2 < 3. The corresponding levels could be voltage, current or charge levels. 
\begin{itemize} 
\item Charge levels. This approach is used in flash memories.  4-valued (MLC) flash memories store two bits per cell. 8-valued (TLC) memories store 3 bits per cell. In 2018,  ADATA, Intel, Micron, and Samsung have launched some SSD products using QLD NAND-memory with 4 bits per cell. While binary flash memories have the advantage of faster write speeds, lower power consumption and higher cell endurance, M-valued flash memories provide higher data density and lower costs. But charges are not suitable for combinational circuits
\item Current levels. Current levels have been used, but are no longer suitable because of the static power dissipation. Power dissipation is the main issue in to-day integrated circuits. 
\item Voltage levels. This is the only practical approach to design combinational circuits.
\end{itemize}

\subsection{Three or one power supplies}

\begin{figure}[htbp]
\centerline{\includegraphics  [width =3 cm]{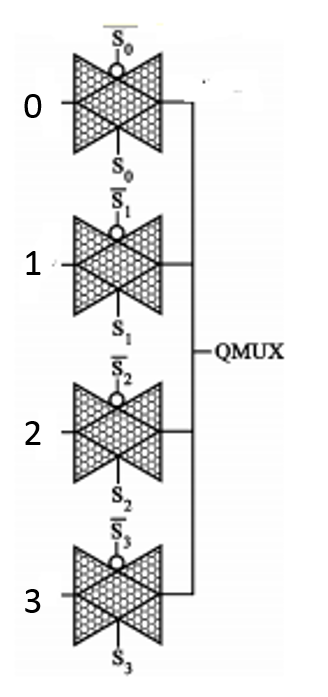}}
\caption{4 voltage levels with 3 power supplies \cite{Moaiyeri}}
\label{43encoder}
\end{figure}

\begin{figure}[htbp]
\centerline{\includegraphics  [width =6 cm]{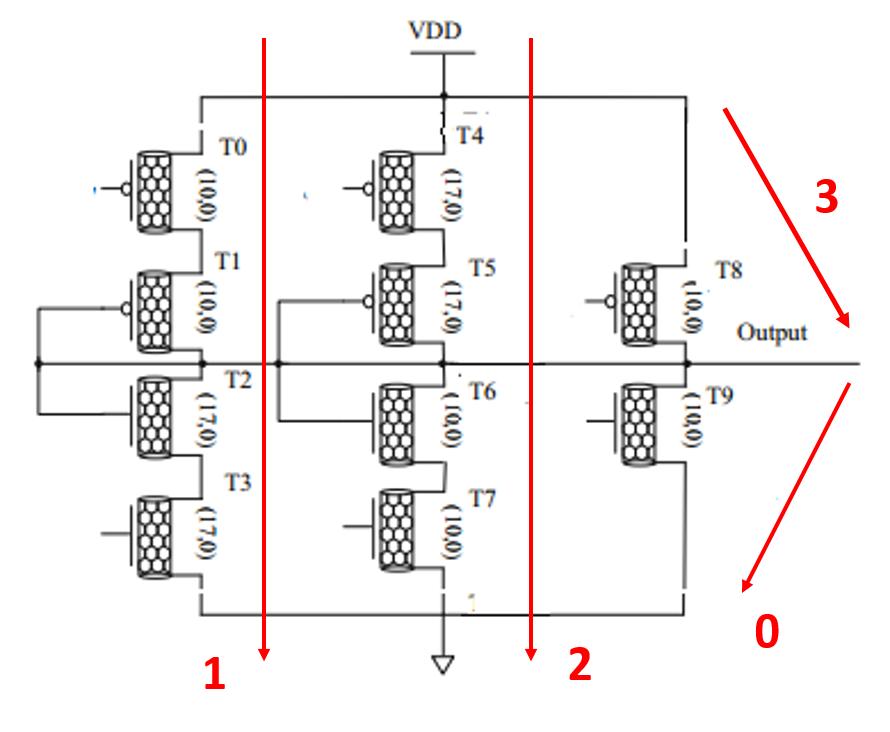}}
\caption{4 voltage levels with 1 power supply \cite{Roosta}}
\label{411encoder}
\end{figure}

\begin{figure}[htbp]
\centerline{\includegraphics  [width =5 cm]{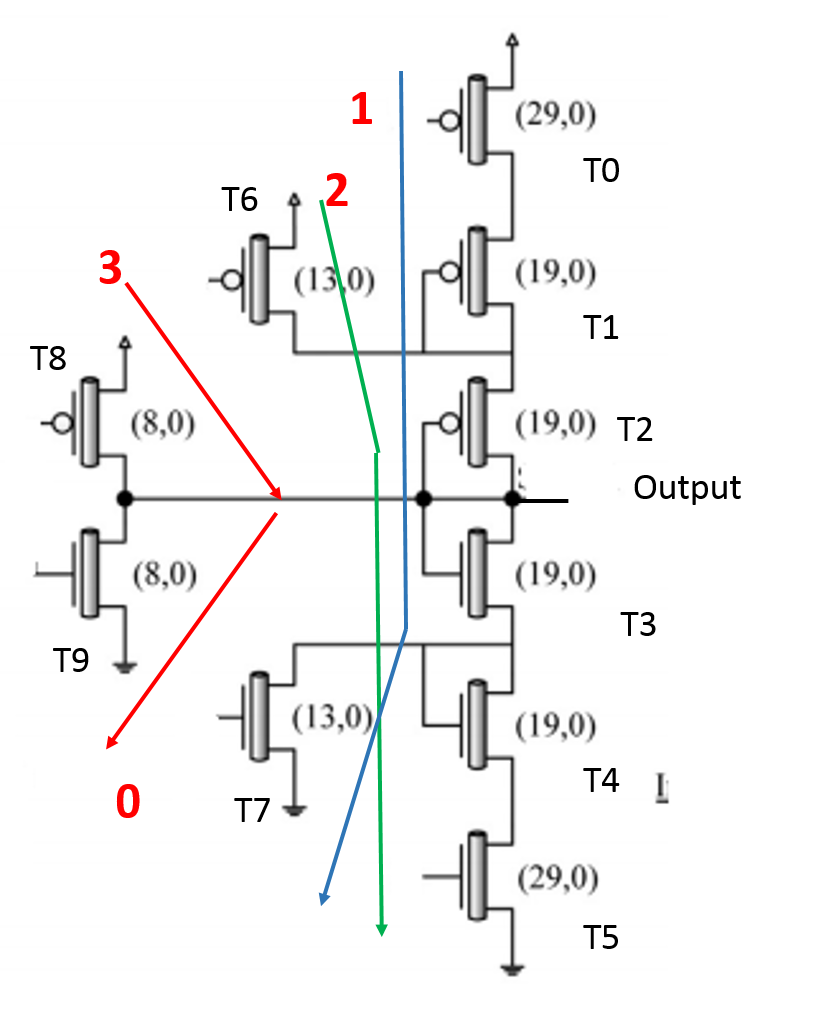}}
\caption{4 voltage levels with 1 power supply \cite{Ebrahimi}}
\label{412encoder}
\end{figure}

The first approach to get four different voltage levels is to use three power supplies:  $ V_{dd}/3$ , $2V_{dd}/3$ and $ V_{dd}$. Fig. \ref{43encoder} presents a possible implementation using transmission gates. $S_{0},  S_{1}, S_{2}, S_{3}$
 true and complementary control inputs are used to transmit to the output one of the four voltage levels corresponding to 0, 1, 2 and 3. The 3 power supplies version of \cite{Roosta} uses the same scheme. This approach drawback is to use three voltage supplies instead of one in the binary case.
The second approach uses only one power supply for levels 0 and 3 and generates levels 1 and 2 through resistor-like dividers.
Fig. \ref{411encoder} shows a first implementation. There are four several pathes: only one should be active to get each output value. Transistors T1, T2, T5, T6 are always on (resistor behavior). The inputs of the other transistors should be fixed to get these transistors on or off.
\begin{itemize} 
\item Level 0 : T9 on ; T0, T3, T4, T7 and T8 off
\item Level 1 : T0 and T3 on ; T4, T7, T8 and T9 off
\item Level 2 : T4 and T7 on ; T0, T3, T8 and T9 off
\item Level 3 : T8 on ; T0, T3, T4, T7 and T9 off
\end{itemize}
Fig. \ref{412encoder} presents a variant of the previous one. Only one path with resistor-like transistors is used with two resistor-connected p and two resistor-connected n transistors. T6  is used to bypass T1 and T7 is used to bypass T4.
\begin{itemize} 
\item Level 0 : T9 on ; T0, T5, T6, T7 and T8 off
\item Level 1 : T0 and T7 on ; T5, T6, T8 and T9 off
\item Level 2 : T5 and T6 on ; T0, T7, T8 and T9 off
\item Level 3 : T8 on ; T0, T5, T6, T7 and T9 off 
\end{itemize}
Both circuits are similar with 10 transistors. This approach has two drawbacks. Levels 1 and 2 generates static power dissipation. The resistors in pathes 1 and 2 increase the RC loads and degrade switching times compared to pathes 0 and 3.

\subsection{Encoder and decoder circuits}
\begin{figure}[htbp]
\centerline{\includegraphics  [width =8 cm]{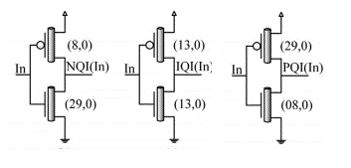}}
\caption{Decoder circuits presented in \cite{Ebrahimi}}
\label{4to2decoder}
\end{figure}

The encoder circuits can be derived from the circuits presented in Fig. \ref{43encoder}, Fig. \ref{411encoder} and Fig. \ref{412encoder}.
The decoder circuits are easy to implement. They correspond to Table \ref{Q2B} in which binary values are 0 and 3. NQI, IQI and PQI outputs are provided by 3 inverters having 3 different threshold levels. Fig. \ref{4to2decoder} shows the corresponding circuits presented in \cite{Ebrahimi}. The situation is similar whether circuits use 3 or 1 power supplies. Appropriate threshold levels are got by defining the chiral number of each transistor used in the inverter.

\begin{table}
\centering
\caption{Truth table of decoder circuits}
\begin{tabular}{|c||c|c|c|}
  \hline
 IN&NQI&IQI&PQI\\
\hline
 0&3&3&3\\
 1&0&3&3\\
 2&0&0&3\\
 3&0&0&0\\
  \hline
\end{tabular}
\label {Q2B}
\end{table}

\section{How to implement a quaternary adder}
Table \ref{T1} shows the truth table of a 1-digit quaternary adder.
There are different techniques to implement a quaternary 1-digit adder:
\begin{itemize}
\item The simplest way is to use a 2-bit binary adder and to interface it with a 4-to-2 decoder and a 2-to-4 encoder. The corresponding adder is presented in section \ref{L1}.
\item The opposite approach is the direct implementation of Table \ref{T1} by using the general approach. A function f(inputs) is decompose into f(inputs) = 3.f3 + 2.f2 + 1.f1 where f3, f2 and f1 are respectively the binary functions of the inputs for which the functions have values 3, 2 and 1. f3, f2 and f1 includes the NQI, IQI and PQI functions of input variables (Table \ref{Q2B}). This approach is used in the adder presented in section \ref{L4}.
\item An intermediate approach uses multiplexers to implement subfunctions that can be derived from Table \ref{T1}. An example of subfunction is the successor function: When A = 1 and Ci = 0 then QS = (B+1) mod. 4. Two adders using this approach are presented in sections \ref{L2} and \ref{L3}.
\end{itemize}
From the 1-digit quaternary adder, N-digit quaternary carry propagate (CPA), carry lookead (CLA) and carry save (CSA) adders can be easily derived.

\begin{table}
\centering
\caption{Truth table of a quaternary adder}
\begin{tabular}{|c|c|c||c|c|c|c|c|c||c|c|}
  \hline
A&B&Ci&QS&QC& &A&B&Ci&QS&QC\\
\hline
 0&0&0&0&0&&0&0&1&1&0\\
0&1&0&1&0&&0&1&1&2&0\\
0&2&0&2&0&&0&2&1&3&0\\
0&3&0&3&0&&0&3&1&0&1\\

 1&0&0&1&0&& 1&0&1&2&0\\
1&1&0&2&0&&1&1&1&3&0\\
1&2&0&3&0&&1&2&1&0&1\\
1&3&0&0&1&&1&3&1&1&1\\

 2&0&0&2&0&& 2&0&1&3&0\\
2&1&0&3&0&&2&1&1&0&1\\
2&2&0&0&1&&2&2&1&1&1\\
2&3&0&1&1&&2&3&0&2&1\\

 3&0&0&3&0&&3&0&1&0&1\\
3&1&0&0&1&&3&1&1&1&1\\
3&2&0&1&1&&3&2&1&2&1\\
3&3&0&3&1&&3&3&1&3&1\\
  \hline
\end{tabular}
\label {T1}
\end{table}

\section{Quaternary adders with quaternary to binary interfaces}
\label{L1}
The simpliest way to implement a quaternary adder is to interface a 2-bit binary adder with quaternary to binary decoder and encoder circuits.
Table \ref{Q2Bconversion} presents the truth table of the quaternary to binary conversion. Binary values are 0 and 3.
\subsection{4 to 2 decoder circuit}
\begin{figure}[htbp]
\centerline{\includegraphics  [width =6 cm]{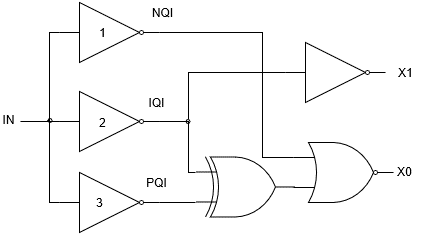}}
\caption{Quaternary to Binary Decoders}
\label{Q2Bdecoder}
\end{figure}

\begin{figure}[htbp]
\centerline{\includegraphics  [width = 8 cm]{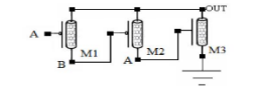}}
\caption{CNTFET 3T Xor}
\label{3Txor}
\end{figure}

The decoder circuit is presented in Fig \ref{Q2Bdecoder}. The circuitry is the same using 3 or 1 voltage levels. It is based on the inverters 1, 2 and 3 with the different threshold levels (such as the inverters presented in Fig. \ref{4to2decoder}) followed by usual binary gates. The number of transistors depends on the implementation of the XOR gate. It ranges from 16 T when using 4 Nand gates down to 3 T as proposed in \cite{nehru} (Fig.\ref{3Txor}). An acceptable value is 9 T, which corresponds to a conventional CMOS implementation used in \cite{xor}. This implementation doesn't use pass transistors and has a full swing output. The overall transistor count for the decoder ranges from 28 T (most conservative implementation) down to 15T with 21 T as an acceptable value.

\begin{table}
\centering
\caption{Truth table of decoder circuits}
\begin{tabular}{|c||c|c|c||c|c|}
  \hline
 Q&NQI&IQI&PQI&X1&X0\\
\hline
 0&3&3&3&0&0\\
 1&0&3&3&0&3\\
2&0&0&3&3&0\\
3&0&0&0&3&3\\
  \hline
\end{tabular}
\label {Q2Bconversion}
\end{table}

\subsection{2 to 4 encoder circuits}
The binary to quaternary encoder circuits depend on the technique that is used to generate the four output values.
\subsubsection {Encoder of Fig. \ref{43encoder}}
\label{E1}
\begin{figure}[htbp]
\centerline{\includegraphics  [width = 4 cm]{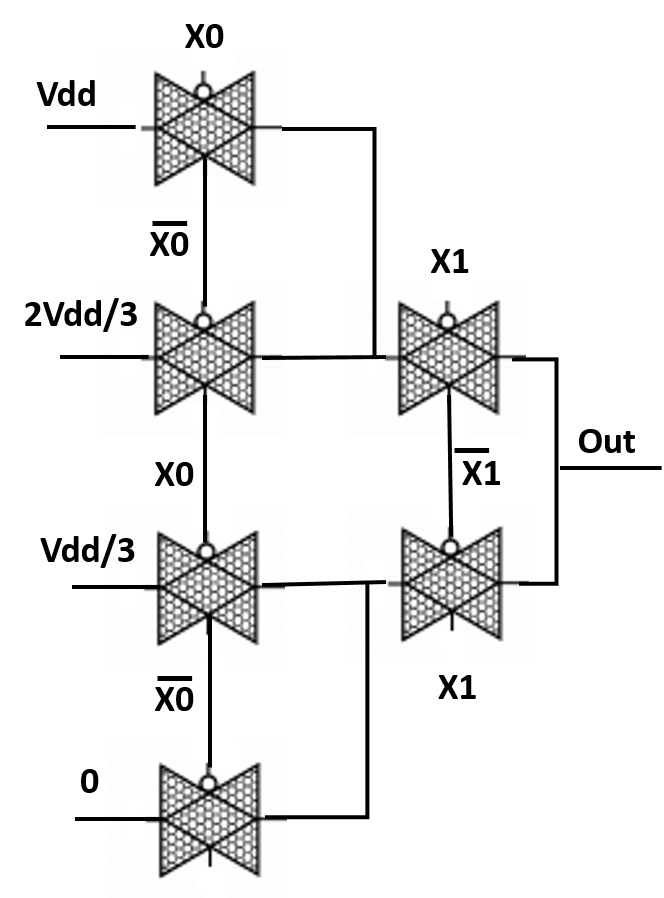}}
\caption{Binary to Quaternary Encoder}
\label{Q2BEncoder}
\end{figure}

The encoder circuit corresponding this approach is shown in Fig. \ref{Q2BEncoder}. It uses 16 T.

\subsubsection {Encoder of  Fig. \ref{411encoder}}
\label{E2}
The inputs of transistors T0, T3, T4, T7, T8 and T9 should be controled. p transistors are on when the input is 0 and n transistors are on when the input is 1.
The corresponding truth table is shown in Table \ref{Tcontrol}.
The corresponding equations are
\begin{itemize}
\item $IT0=\overline{\overline{X1}.X0}= NAND(\overline{X1},X0)$ 
\item$ IT3 = NOT(IT1)$
\item $IT4 = \overline{X1.\overline{X0}}= NAND(X0).\overline{X1}$
\item$ IT7=NOT(IT4)$
\item $IT8=\overline{X1}+\overline{X0}=NAND(X1,X0)$
\item $IT9 = \overline{X1}.\overline{X0} = NOR(X1,X0)$
\end{itemize}

4 NOT gates are needed ($\overline{X0}, \overline{X1}$, IT3 and IT4), together with 3 Nand and 1 Nor gates to control the inputs. The total transistor count is 8 (NOT) + 16 (Nand and Nor) + 10 (Fig. \ref{411encoder}) = 34 T.

\begin{table}
\centering
\caption{Controling transistors in Fig. \ref{411encoder}}
\begin{tabular}{|c|c||c|c|c|c|c|c|}
  \hline
X1&X0&IT0&IT3&IT4&IT7&IT8&IT9\\
\hline
 0&0&1&0&1&0&1&1\\
0&1&0&1&1&0&1&0\\
1&0&0&1&1&1&1&0\\
1&1&1&0&1&0&0&0\\
  \hline
\end{tabular}
\label {Tcontrol}
\end{table}

\subsubsection {Encoder of  Fig. \ref{412encoder}}
\label{E3}
The inputs of transistors T0, T5, T6, T7, T8 and T9 should be controled. p transistors are on when the input is 0 and n transistors are on when the input is 1.
The corresponding truth table is shown in Table \ref{Tcontrol2}.
The corresponding equations are
\begin{itemize}
\item $IT0=\overline{\overline{X1}.X0}= NAND(\overline{X1},X0)$ 
\item $IT5=\overline{X1.\overline{X0}}= NOR(\overline{X1},X0)$ 
\item $IT6 = NOT(IT5)$
\item$ IT7=NOT(IT0)$
\item $IT8=\overline{X1}+\overline{X0}=NAND(X1,X0)$
\item $IT9 = \overline{X1}.\overline{X0} = NOR(X1,X0)$
\end{itemize}

4 NOT gates are needed ($\overline{X0}, \overline{X1}$, IT6 and IT7), together with 2 Nand and 2 Nor gates to control the inputs. The total transistor count is 8 (NOT) + 16 (Nand and Nor) + 10 (Fig. \ref{412encoder}) = 34 T.

\begin{table}
\centering
\caption{Controling transistors in Fig. \ref{412encoder}}
\begin{tabular}{|c|c||c|c|c|c|c|c|}
  \hline
X1&X0&IT0&IT5&IT6&IT7&IT8&IT9\\
\hline
 0&0&1&0&1&0&1&1\\
0&1&0&0&1&1&1&0\\
1&0&1&1&0&0&1&0\\
1&1&1&0&1&0&0&0\\
  \hline
\end{tabular}
\label {Tcontrol2}
\end{table}

\subsubsection {Transistor count for encoder and decoder circuits for quaternary to binary interfaces}
The transistor count is
\begin{itemize}
\item 28/15 (decoder) + 16 (encoder) = 44/31 T for the first implementation (subsection \ref{E1}), according to the implementation of the Xor gate. The transistor count is the same for the third implementation with 3 supply voltages (subsection \ref{E3}).
\item 28/15 (decoder) + 34 (encoder) = 62/49 T for the two subsequent implementations (subsections \ref{E2} and \ref{E3}) with a single-supply.
\end{itemize}
\subsection{1-digit quaternary adder using a binary adder}

\begin{figure}[htbp]
\centerline{\includegraphics  [width = 8 cm]{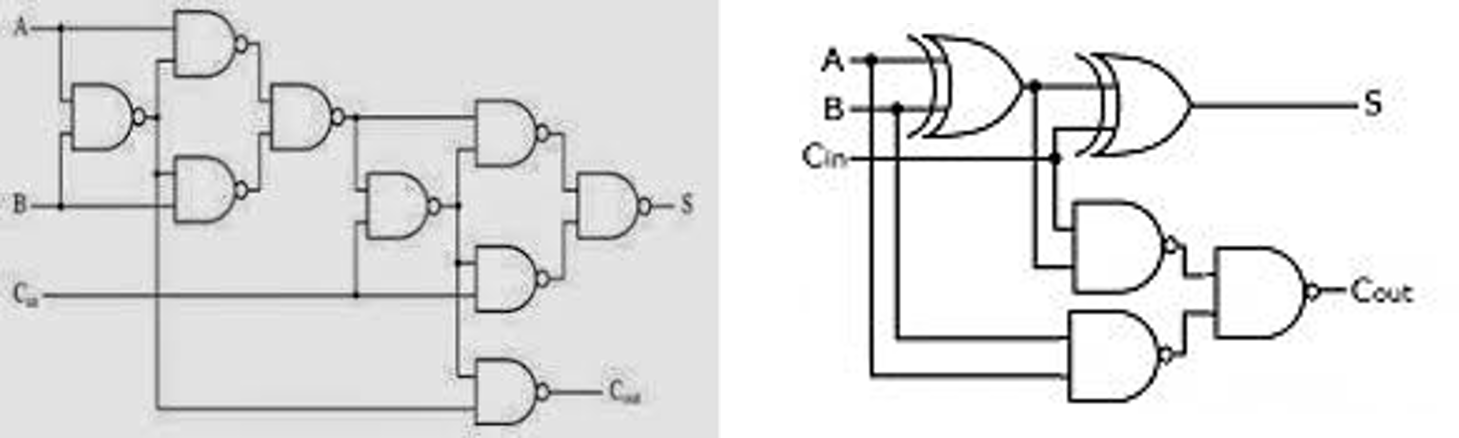}}
\caption{Binary full adders}
\label{FA}
\end{figure}

\begin{figure}[htbp]
\centerline{\includegraphics  [width = 9 cm]{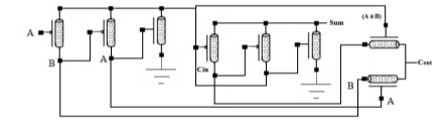}}
\caption{8T binary full adder}
\label{8TFA}
\end{figure}

There are many different ways to implement binary adders. They differ on the use or not of transmission gates. It is out of the scope of this paper to present all the possible implementations. 
Fig. \ref{FA} presents two typical implementations of a full adder. The left part only uses Nand gates. The right part uses Xor and Nand gates. A CNTFET 8 T full adder (Fig. \ref{8TFA}) has been presented \cite{nehru}. This adder doesn't restore levels and using it could raise issues, both for noise margins and switching times due to series of pass transistors. The transistor counts are respectively 36 T, 18 T and 8 T.
The quaternary adder uses two binary adders, one encoder and one decoder circuits. 
Using 2-bit carry propagate adders, the overall transistor count for the 3 power supplies version is thus:
\begin{itemize}
\item 72 + 44 = 116 T without using pass transistors
\item 36 + 31 = 67 T when using pass transistors for Xor gates
\item 16 + 31 = 47 T when using pass transistors for Xor gates and the 8T binary adder (Fig. \ref{8TFA})
\end{itemize}
The single-supply version would use more transistors (+ 18 T).

\subsection{N-digit quaternary adders}
Using quaternary interfaces and 2N-bit adders, N-digit quaternary adders can be implemented. CPAs, CLAs and CSAs implementations are discussed in section \ref{Carry}.

\section{Quaternary adders presented in \cite{Ebrahimi}}
\label{L4}
These adders are based on the following approach: $Qs(inputs) = 3.f3(inputs) + 2.f2(inputs) + 1.f1(inputs)$ where fi(inputs) is the binary function for which Qs = i. Any input must be decomposed according to Table \ref{T3}. The corresponding circuit is shown in Fig. \ref{F1LIFL}. It uses 18 T.
For the half adder, according to the left part of Table \ref{T1}, the equations are \\
$Sum = 3.(A0.B3+A1.B2+A2.B1+A3.B0)\\ 
\hspace* {1 cm}+2.(A0.B2+A1.B1+A2.B0+A3.B3)\\
\hspace* {1 cm}+1.(A0.B1+A1.B0+A2.B3+A3.B2)\\
Carry = 1.(A1.B3+A2.B2+A2.B3+A3.B1.A3.B2\\
\hspace* {1 cm}+A3.B3)$

The half adder circuit is presented in Fig. \ref{F2LIFL}. With 2 input decoders, the sum circuit and the carry circuit, the transistor count is 87 T.
The corresponding full adder presented in \cite{Ebrahimi} has a quaternary carry input. While this could be useful for designing compressors used in multiplier reduction trees, it is useless for usual N-digit adder in which carry input and output have binary values.
 In Fig. {\ref{F4LIFL}, we present a modified version in which binary carries are used. 
The half adder implements the H function, defined as H = (A+B) mod 4. A modified half adder implements Sum = H + C. It has the decoded values of quaternary input H (provided by the Q-dec shown in Fig. \ref{F1LIFL}) and the binary carry input. The corresponding scheme is shown in Fig. \ref{MFASUM}. With one Q-Dec (H), one NQI inverter + one binary inverter to generate C0 and C1, it has 8 T + 4 T + 28 T = 40 T while the sum part of the quaternary half-adder has 52 T.
The carry generator circuit is based on the following observations:
\begin{itemize}
\item $0 \leq A+B+Cin \leq 7$
\item Cout=0 iff  A+B+Cin<4  and Cout = 1 iff A+B+Cin>3
\item Cout=0 if (Cin= 0 and A+B<4) or (Cin=1 and A+B < 3)
\item The correspondance between H = (A+B) mod. 4 and A+B is given in Table \ref{T4}
\item From Table \ref{T4}, $\overline{ Cout}= H0.A0+ H1. Ai+H2.A3+\overline{Cin}.\overline{H3})$
\end{itemize}
The corresponding carry generator circuit is shown in Fig. \ref{MFACAR2}.

The complete modified quaternary adder has 52 T (sum part of QHA) + 40 T (sum part of modified QHA) + 19 T (carry circuit) = 111 T. This number is minimal, as the minimal number of Q-DEC is used, assuming that there are no fan-out or routing issues.

\begin{table}
\centering
\caption{Decoding of quaternary inputs \cite{Ebrahimi}}
\begin{tabular}{|c||c|c|c|c|c|c|c|c||c|c|}
  \hline
I&I0&I1&$\overline{I1}$&Ii&$\overline{l2}$&I2&I3\\
  \hline
0&3&0&3&3&3&0&3\\
1&0&3&0&3&3&0&3\\
2&0&0&3&0&0&3&3\\
3&0&0&3&0&3&0&0\\

  \hline
\end{tabular}
\label {T3}
\end{table}

\begin{table}
\centering
\caption{Carry out computation}
\begin{tabular}{|c|c|c|c||c||c|c|c|c|c|c}
  \hline
Cin&A+B&H&Cout&&Cin&A+B&H&Cout\\
  \hline
0&0&0&0&&1&0&0&0\\
0&1&1&0&&1&1&1&0\\
0&2&2&0&&1&2&2&0\\
0&3&3&0&&1&3&3&1\\
0&4&0&1&&1&4&0&1\\
0&5&1&1&&1&5&1&1\\
0&6&2&1&&1&6&2&1\\

  \hline
\end{tabular}
\label {T4}
\end{table}

\begin{figure}[htbp]
\centerline{\includegraphics  [width =8 cm]{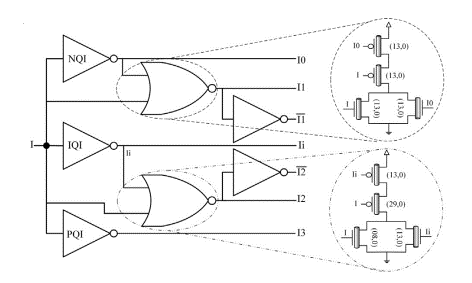}}
\caption{Complete quaternary decoder presented in \cite{Ebrahimi}} 
\label{F1LIFL}
\end{figure}

\begin{figure}[htbp]
\centerline{\includegraphics  [width =9 cm]{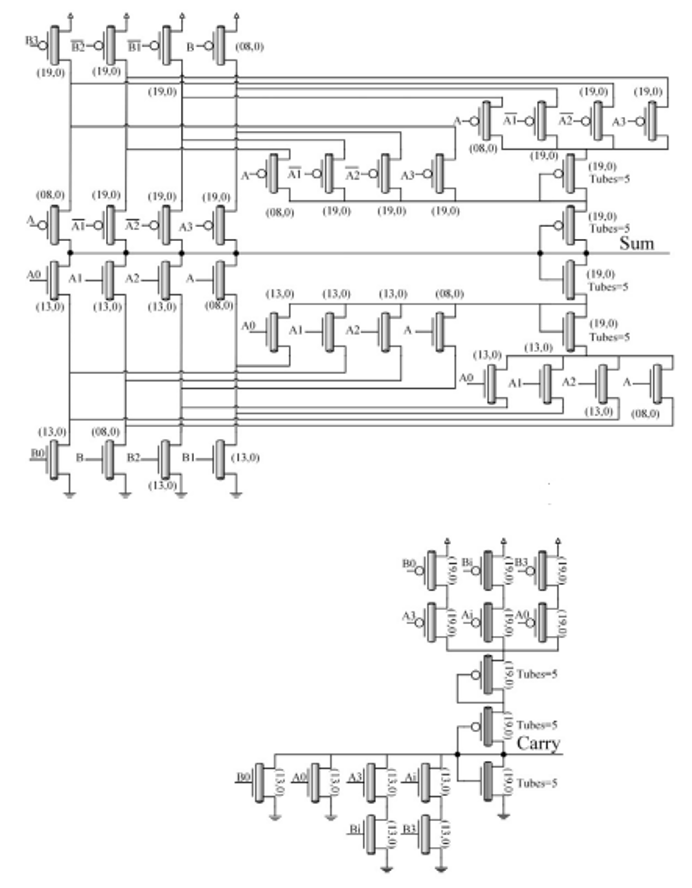}}
\caption{Half adder presented in \cite{Ebrahimi}} 
\label{F2LIFL}
\end{figure}

\begin{figure}[htbp]
\centerline{\includegraphics  [width =4 cm]{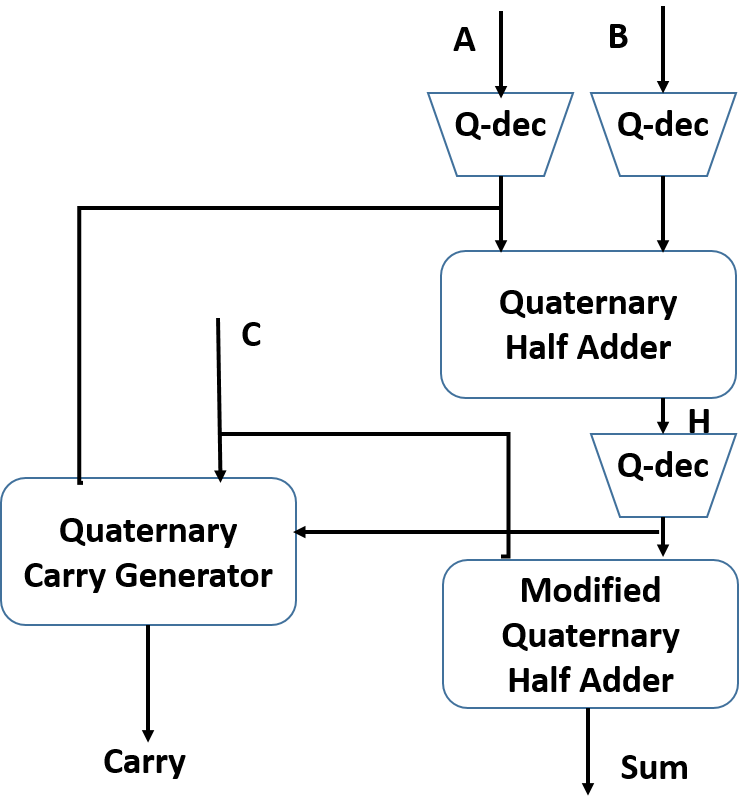}}
\caption{Modified full adder from \cite{Ebrahimi}} 
\label{F4LIFL}
\end{figure}

\begin{figure}[htbp]
\centerline{\includegraphics  [width =4 cm]{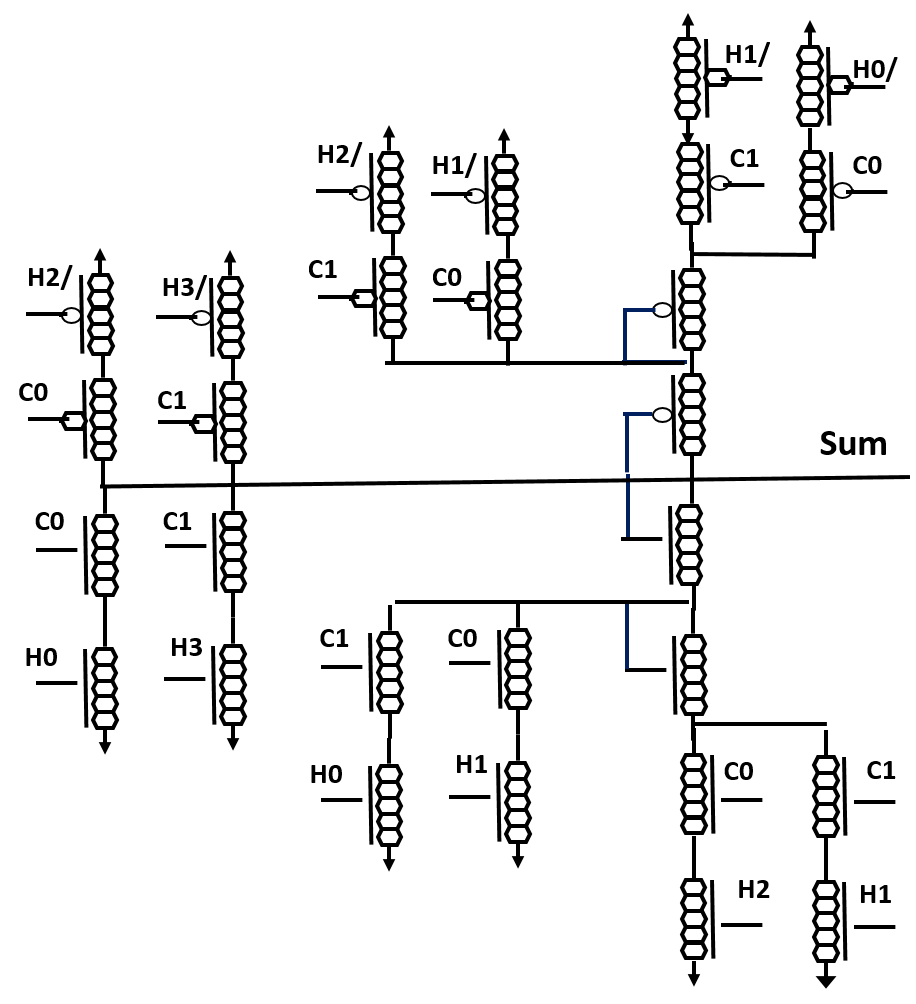}}
\caption{Modified full adder- Sum circuit} 
\label{MFASUM}
\end{figure}

\begin{figure}[htbp]
\centerline{\includegraphics  [width =4 cm]{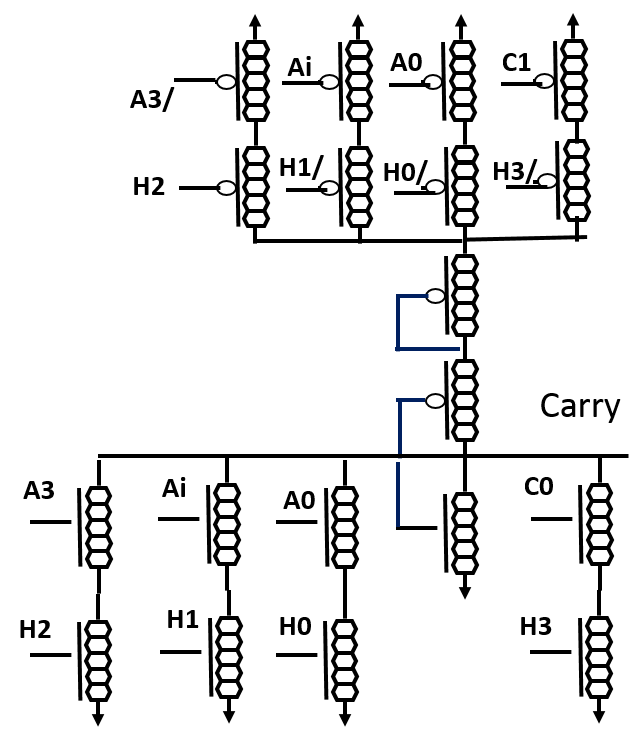}}
\caption{Modified full adder- Carry circuit} 
\label{MFACAR2}
\end{figure}

\section{MUX based quaternary adders}

The MUX based implementation is based on the observation of the quaternary half adder truth table (left part of Table \ref{T1} when Ci = 0). When A = 0 then QS = B. When A=1 then QS = (B+1) mod 4 (successor B). When A=2, QS = (B+2) mod 4 (2nd level successor B). When B=3 then QS = (B-1) mod. 4 (predecessor B).
\subsection{Quaternary adder derived from \cite{Moaiyeri}} 
\label{L2}
The quaternary half adder presented in \cite{Moaiyeri} uses the decoder circuits and the muxes presented in Fig. \ref{MF1}. The QTG circuits are used to implement the successor and predecessor functions. Transistor counts for QDEC and QMUX are both 16 T. The half adder based on QDEC and QMUX is presented in Fig.\ref{MF2}. The transistor counts are
\begin{itemize}
\item For QS, there are 4 QTGs and 2 QDECs for a total of 16 T*6 = 96 T.
\item For QCarry, there are 6 inverters, 6 transistors and 1 QTG for a total of 12 + 6 + 16 = 32  T.
\item The half adder has 128 T.
\end{itemize}
The corresponding full adder is not presented in \cite{Moaiyeri}. However, the full adder can be easily derived. The half adder (Fig. \ref{MF2}) corresponds to C=0. To compute QSUM1 corresponding to C=1, only two more QTGs are needed. The final sum is derived from QSUM0 and QSUM1 by using two transmission gates and one inverter. A similar technique is used to compute the carry output, as shown in Fig. \ref{MF4}. Only one more QTG and two transmission gates are needed.
The overall transistor count for the full adder is 96 + 32 + 6 + 16 + 4 = 154 T.

\begin{figure}[htbp]
\centerline{\includegraphics  [width = 9 cm]{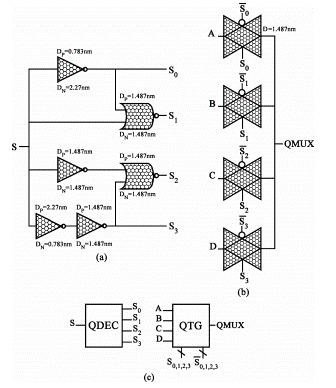}}
\caption{Decoders and Muxes from \cite{Moaiyeri}}
\label{MF1}
\end{figure}

\begin{figure}[htbp]
\centerline{\includegraphics  [width = 9 cm]{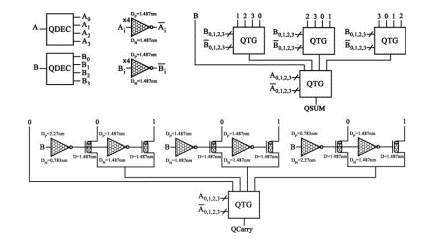}}
\caption{Half adder presented in \cite{Moaiyeri}}
\label{MF2}
\end{figure}

\begin{figure}[htbp]
\centerline{\includegraphics  [width = 9 cm]{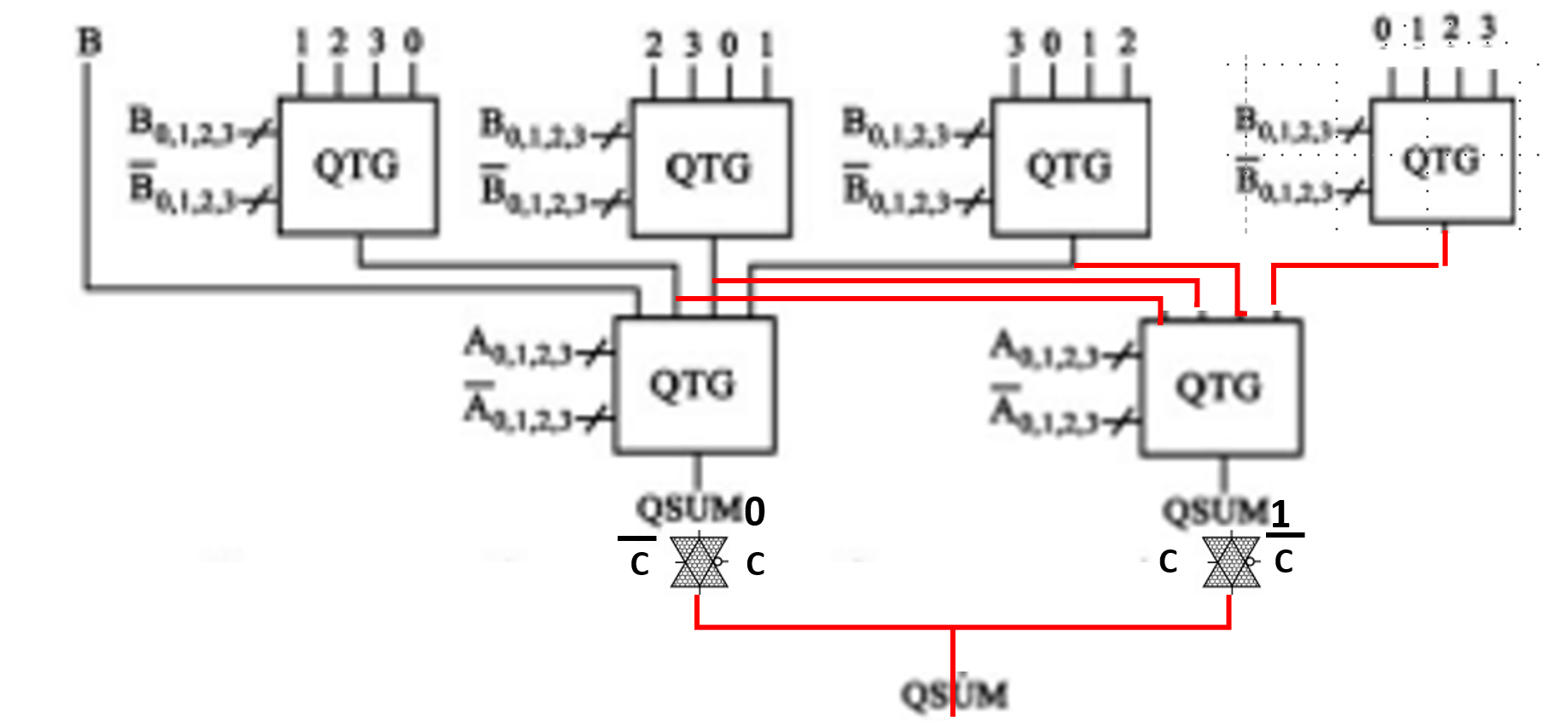}}
\caption{Full adder sum output derived from \cite{Moaiyeri}}
\label{MF3}
\end{figure}

\begin{figure}[htbp]
\centerline{\includegraphics  [width = 9 cm]{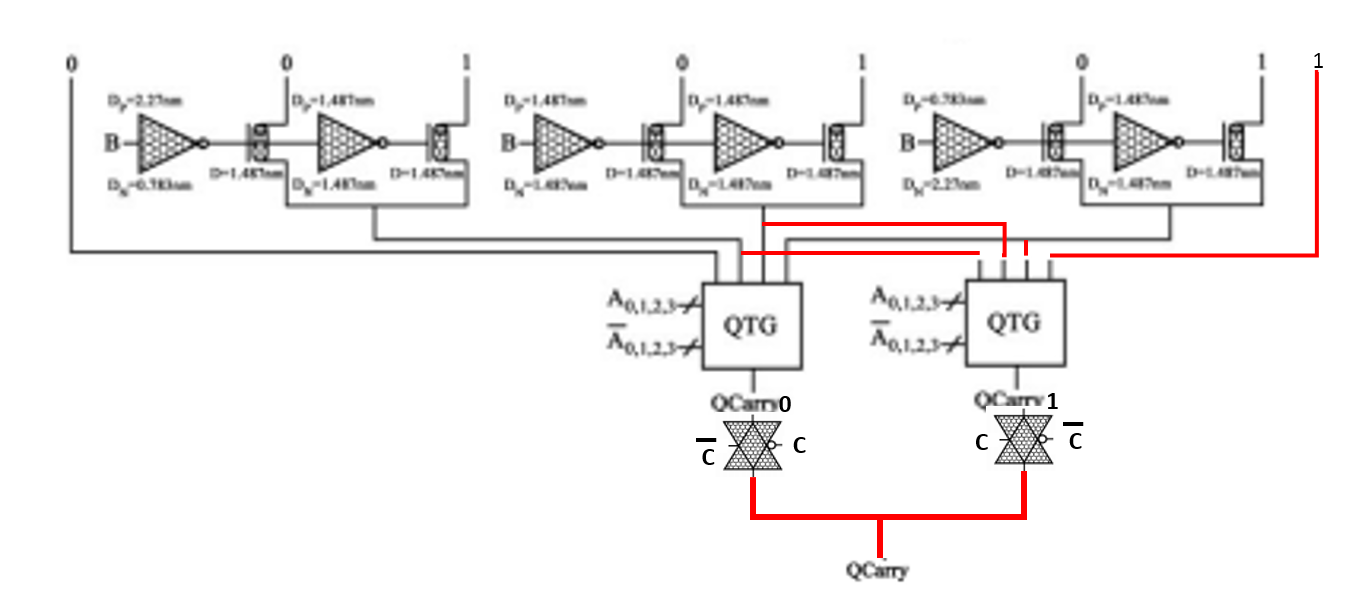}}
\caption{Full adder carry output derived from \cite{Moaiyeri}}
\label{MF4}
\end{figure}

\subsection{Quaternary adders presented in \cite{Roosta}}
\label{L3}
These adders also use MUXes, but implement the successor, second level successor and predecessor circuits as separate blocks. Basically, the half adder presented in Fig. \ref{RF1} is similar to the half adder of Fig. \ref{MF2}. The corresponding full adder, presented in Fig. \ref{RF2}, also use the same approach than the full adder of Fig. \ref {MF3} and Fig. \ref{MF4}. Two versions are presented, with one and three power supplies. The different components are
\begin{itemize}
\item QMUX 4:1 is shown in Fig. \ref{RF3}. It has 12 T.
\item QMUX (not shown) is simplier with only 6T.
\item The successor circuit with 3 power supplies is shown in Fig. \ref{RF4}. It has 6 T. The second level successor  predecessor circuits (not shown) have also 6 T. The transistor count for the 3 circuits is 18 T.
\item The successor circuit with 1 power supply is shown in Fig. \ref{RF5}. It has 13  T. The second level successor and the predecessor circuits (not shown) have respectively 12 T and 17 T.  The transistor count for the 3 circuits is 42 T.
\item Inverters are needed for $\overline{NQI(B)},  \overline{IQI(B)},  \overline{PQI(B)}$,  $\overline{NQI(S\_QHA)},  \overline{IQI(S\_QHA)}$, $\overline{PQI(S\_QHA)}$. 
\begin{itemize}
\item 3 power supplies: If the  B inverters drive the different subblocks, the fan-out are respectively  10, 6 and 8. Only 3 inverters (6 T) are needed, but there could be fan-out and routing issues. If  different B inverters are used for each subblocks, there are 12 inverters (24 T). 
\item 1 power supply: If the B inverters drives the different subblocks, the fan-out are respectively 11, 9 and 11. There are 3 inverters (6 T). With different inverters for each subblock, there are 12 inverters (24 T).
\end{itemize}  
\end{itemize}

The overall transistor count is given in Table \ref{T2}. Obviously, the 3 power supplies version is more efficient than the version presented in  \cite{Moaiyeri}: customizing the implementation of the successor and predecessor functions reduces the transistor count versus using 4-valued MUXes. The 1-power supply version has far more transistors. 
\begin{figure}[htbp]
\centerline{\includegraphics  [width = 9 cm]{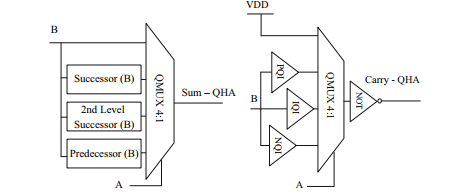}}
\caption{Half adder presented in \cite{Roosta}}
\label{RF1}
\end{figure}

\begin{figure}[htbp]
\centerline{\includegraphics  [width = 9 cm]{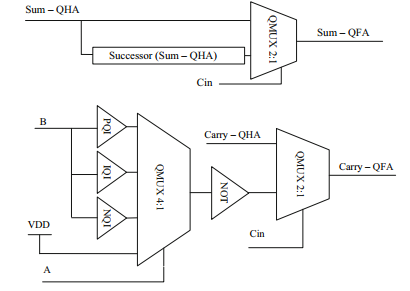}}
\caption{Full adder presented in \cite{Roosta}}
\label{RF2}
\end{figure}

\begin{figure}[htbp]
\centerline{\includegraphics  [width = 9 cm]{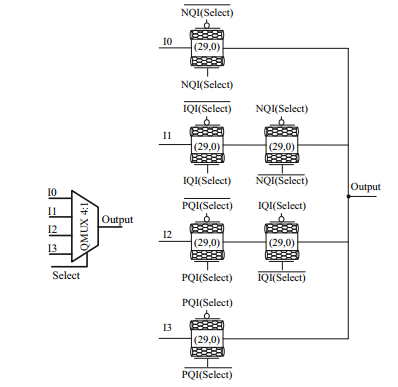}}
\caption{QMUX 4:1 presented in \cite{Roosta}} 
\label{RF3}
\end{figure}

\begin{figure}[htbp]
\centerline{\includegraphics  [width = 6 cm]{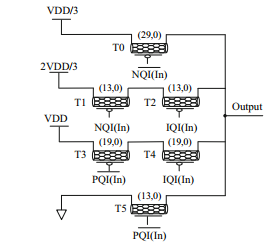}}
\caption{Three voltages successor curcuit presented in \cite{Roosta}} 
\label{RF4}
\end{figure}

\begin{figure}[htbp]
\centerline{\includegraphics  [width = 6 cm]{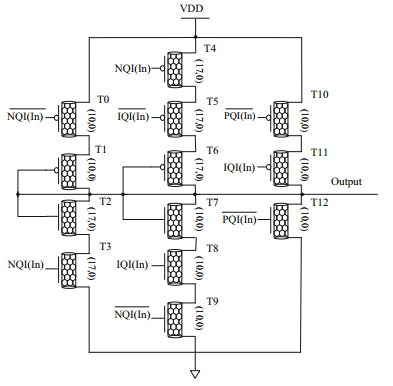}}
\caption{Single power supply successor circuit \cite{Roosta}} 
\label{RF5}
\end{figure}

\begin{table}
\centering
\caption{Transistor count for quaternary adder \cite{Roosta}}
\begin{tabular}{|c|c|c||c|c|c|c|c|c||c|c|}
  \hline
&S-HA&C-HA&SFA&CFA& Inverters&Total\\
\hline
3  supplies&30&14&12&20&6/24&82/100\\
1  supply&54&14&36&20&6/24& 130/148\\

  \hline
\end{tabular}
\label {T2}
\end{table}

\section{Carry Look Ahead and Carry Skip Adders}
\label{Carry}
We now compare the carry computation for a 8-bit and 4-digit CLA and CSA adders. The binary computation is decomposed in two 4-bit blocks. The quaternary computation only uses one block.
\subsection{Carry-Look Ahead Adders}
Fig. \ref{4CLA} presents a 4-bit carry look-ahead adder. The binary equations of the carry computation part are well-known: \\
$Gi=Ai.Bi$ \\
$Pi = Ai \oplus  Bi$ (or $Pi=Ai+Bi)$\\
$C1= G0 +P0.C0$\\
$C2 = G1 + G0.P1 + P0.P1.C0= G2 + P1(G0+P0C0)$\\
$C3 = G2 + G1.P2 + G0.P1.P2 + P0.P1.P2.C0 = G2 + G1P2+P2P1(G0+P0C0)$\\
$C4 = G3 + G2.P3 + G1.P2.P3 + P1.P2.P3(G0+ P0C0)$ \\
Binary Gi and Pi functions are implemented respectively by Nand + Inverter and Nor + inverter. Both functions use 6 T. 
The optimal implementation of C1, C2, C3 and C4 uses a complex gate + one inverter.  The transistor count for a 4-bit carry computation is given in Table \ref{BCCLA}.
For quaternary adders, the binary G and P functions for any bit j are:\\
$G=((A=3)\wedge( B \geq1))+((A\geq2) \wedge (B\geq2))+ \\
 \hspace* {1 cm}((B=3)\wedge (A \geq 1))$\\
 $P =A3.B1+A2.B2+A1.B3$

According to Table \ref{T3}, the equations can be reformulated as\\
$G = \overline{(A3+B0).(Ai+Bi).(B3+A0)}$\\
$P = \overline{\overline{A3.B1}.\overline{A2.B2}.\overline{A1.B3}}$ \\
where A0 and B0 are the outputs of NQI inverters, Ai and BI are the outputs of IQI inverters, A3 and B3 are the outputs of PQI inverters and A1, A2, B1 and B2 are the outputs of the circuit shown in Fig. \ref{F1LIFL}.
Assuming that all these values are available, the transistor count is 12 T for G and 16 T for P.
For 4  digits, the equations are similar with different implementations of Gi and Pi functions. The transistor count for a 4-digit carry computation is given in Table \ref{QCCLA}.

\begin{table}[!b]
\caption{Transistor count for the carry computations of a 8-bit CLA}
\label{BCCLA}
\begin{tabular}{|c|c|c|c|c|c|c|c|c|}
\hline
Function&Gi&Pi&C1&C2&C3&C4&4-bit&8-bit\\
\hline
T. count  & 24 &24&8&12&16&20&104 &208\\

\hline
\end{tabular}
\end{table}

\begin{table}[!b]
\caption{Transistor count for the carry computations of a 4 digit CLA Quaternary Adder}
\label{QCCLA}
\begin{tabular}{|c|c|c|c|c|c|c|c|c|}
\hline
Function&Gi&Pi&C1&C2&C3&C4&4 quaternary digits\\
\hline
T. count  & 48 &64&8&12&26&20&168\\

\hline
\end{tabular}
\end{table}

The transistor count is better for the carry computation of quaternary adders versus binary ones. The increase cost of Gi and Pi implementation is compensated by the reduced number of logical levels.

\subsection{Carry-Skip Adders}
For an 8-bit CSA, the binary carry computation is composed of two 4-bit skip computations. For 4-bit, it means P1 to P4 functions, a 4-input And gate and a multiplexer. For a 4-digit CSA, the carry computation uses the same number of functions with the only difference in the implemention of Pi. The transistor counts are given in Table \ref{CCSA}. 

\begin{table}[!b]
\caption{Transistor count for the carry computations of 8-bit and 4-digit CSAs}
\label{CCSA}
\begin{tabular}{|c|c|c|c|c|c|c|c|c|}
\hline
&Pi&Nand+inverter&Mux&4-bit CS&8-bit 4-digit CS\\
\hline
B&24&10&14&48&96\\
\hline
Q &64&10&14&&88 \\
\hline
\end{tabular}
\end{table}

\begin{figure}[htbp]
\centerline{\includegraphics  [width =8 cm]{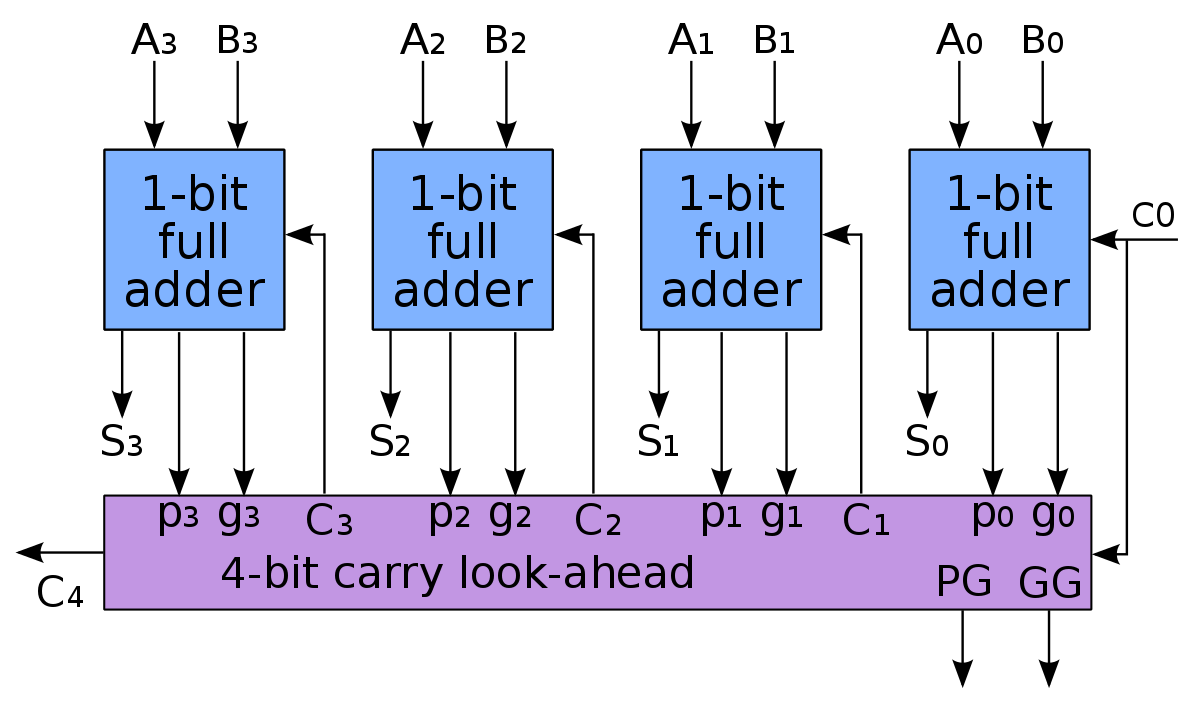}}
\caption{: A 4-bit binary carry look-ahead adder}
\label{4CLA}
\end{figure}

\begin{figure}[htbp]
\centerline{\includegraphics  [width =4 cm]{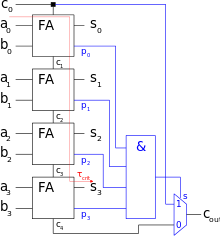}}
\caption{: A 4-bit binary carry skip adder}
\label{4CSA}
\end{figure}

\section{Comparing the different quaternary adders with  binary adders}
\subsection{1-digit quaternary adder versus 2-bit binary adder}
Table \ref{T6} summarizes the transistor count for the different quaternary adders:
\begin{itemize}
\item QB adder corresponds to the binary implementation with binary to quaternary interfaces (section \ref{L1}). The different values correspond to the different ways to implement a binary full adder. The middle value is probably the most significant.
\item QFA  \cite{Ebrahimi} is the adder that was detailed in section \ref{L4}.
\item QFA  \cite{Moaiyeri} is the adder that was detailed in section \ref{L2}.
\item QFA \cite{Roosta} is the adder that was detailed in section \ref{L3}.
\end{itemize}
With one power supply, interfacing a 2-bit adder with quaternary to binary interface is the best implementation. With 3 power supplies, there is no significant difference with the best MUX quaternary implementation. In both cases, the different quaternary adders have x2 or x3 the transistor count of a typical 2-bit binary adder.

\begin{table}[!b]
\caption{Transistor count for 1-digit quaternary adders}
\label{T6}
\begin{tabular}{|c|c|c|c|c|c|c|c|}
\hline
P. Supply&QB adder& QFA \cite{Ebrahimi}& QFA \cite{Moaiyeri}&QFA \cite{Roosta} &2-bit FA\\
\hline
1&134/\textbf{85}/65&111&&148/130&72/\textbf{36}/16\\
\hline
3 &116/\textbf{67}/47& &154&100/82 &\\
\hline
\end{tabular}
\end{table}

\subsection{4-digit quaternary adders versus 8-bit binary adders}
Table \ref{T7} and Table \ref{T8} summarize the transistor count for the different implementations of 4-digit quaternary adders to be compared with a 8-bit binary adder. Within these tables, 
\begin{itemize}
\item First column is the adder type.
\item Second column is the quaternary adders built from a 8-bit binary adder with 4-to-2 decoders and 2-4 encoders. The three values correspond to 1) implementation without pass transistor, 2) a conventional implementation with pass transistors and 3) a debatable option where the Xor implementation could raise noise and switching issues. The second value is the most trustable one. 
\item Third column in Table \ref{T7} corresponds to the straigthforward implementation according to the quaternary functions using 1 power supply. 
\item Fourth column  in Table \ref{T8} corresponds to quaternary adders (3 power supplies) using Muxes.
\item Fifth column corresponds to implementations with Muxes and customized successor and predecessor circuits.
\item  The last column presents the transistor count for the binary implementation. While this implementation only uses one power supply, it is included to Table \ref{T8} for the comparisons.
\end{itemize}

\begin{table}[!b]
\caption{T. count for 4-digit quaternary adders - 1 power supply}
\label{T7}
\begin{tabular}{|c|c|c|c|c|c|c|c|c|}
\hline
&QB adders&\cite{Ebrahimi} adder&\cite{Moaiyeri} adder&\cite{Roosta} adder&8-bit adder\\
\hline
CPA&536/\textbf{340}/260&444&&592/520&288/\textbf{144}/64\\
CLA&784/\textbf{588}/508&612&&760/688&496/\textbf{352}/272\\
CSA&632/\textbf{436}/356&532&&680/608&384/\textbf{240}/160\\
\hline
\end{tabular}
\end{table}

\begin{table}[!b]
\caption{T. count for 4-digit quaternary adders -  3 power supplies}
\label{T8}
\begin{tabular}{|c|c|c|c|c|c|c|c|c|}
\hline
&QB adders&\cite{Ebrahimi} adder&\cite{Moaiyeri} adder&\cite{Roosta} adder&8-bit adder\\
\hline
CPA&464/\textbf{268}/188&&616&400/328&288/\textbf{144}/64\\
CLA&672/\textbf{476}/396&&784&568/496&496/\textbf{352}/272\\
CSA&560/\textbf{436}/284&&704&488/416&384/\textbf{240}/160\\
\hline
\end{tabular}
\end{table}

Some significant results can be derived from Table \ref{T7} and Table \ref{T8}.
\begin{itemize}
\item With only one power supply, the direct interfacing of a binary adder with 4-2 decoders and 2-4 encoders is the best implementation with the smallest transistor count.
\item With three power supplies, only the implementation proposed in \cite{Roosta} can compete with the interfacing of binary adders. We can notice than the transistor count for this implementation is optimistic as it implies that the minimal number of NQI, IQI and PQI inverters can be used without fan-out and connection issues. All the other implementations are outperformed by the direct interfacing of binary adders.
\item Obviously, the best quaternary adder is outperformed by the binary adder computing the same amount of information. This binary adder is included in the best quaternary adder, while the interfacing decoder and encoder circuits are a significant overhead.
\end{itemize}
Quaternary adders are specific combinational circuits. They have some drawbacks. Either they use three power supplies instead of one for binary circuits, or they exhibit static power dissipation and degraded switching times when using only one power supply. 
However, the main point is that the best implementation of quaternary adders consists in interfacing binary adders with 4 to 2 decoder and 2 to 4 encoder circuits. It means that there is no advantage to try to directly implement quaternary combinational functions. To summarize, the best quaternary adder with N digits is the corresponding 2N binary adder with a significant overhead: decoder and encoder circuits.

\section{Concluding remarks}
Most presented implementations of ternary or quaternary circuits claim advantages of multiple valued circuits. The following quote summarizes the arguments  that may be found in most MVL papers : 
``MVL circuits have potential advantages. Using MVL circuits reduces the complexity of interconnection via reducing the number
of wires since each wire carries more than one digit of data. Power consumption and area of the MVL circuits are generally less than the corresponding binary circuits due to the reduction in number of active elements [8].

How does our results fit with these claims ? It is obvious that a N digit quaternary adder has less input and output digits than a 2N bit binary adder. But we have shown that the best N-digit quaternary adder includes the corresponding 2N bit binary adder with the overhead of input decoder and output encoder circuits. According to Table \ref{T7} and Table \ref{T8}, the best 4-digit quaternary adder has more than 2.5x the transistor count of 8-bit binary adders. These transistor must be interconnected: it means that the quaternary adders have far more connections than the binary adders as soon as the internal connections are considered.
As a matter of facts, is there an ``interconnection wall" in digital circuits as the well-known ``power wall" and a ``memory wall"?. 
The answer is no, even in there could be interconnection isssues in circuits such as FPGAs.
While the up-to-date CMOS technological nodes are more and more costly, they have more and more interconnection layers. Twenty years ago, the 180 nm node had 6 metal layers. To-day, the number of metal layers in nano-CMOS technologies usually ranges from 8 to 15, with a trade-off between integration and cost.

It is difficult to believe that x2.5 more transistors could lead to a reduction of chip area and power dissipation. More transistors means more chip area and more power dissipation. It turns out that the assumptions of the quote are false, at least for using MVL techniques for combinational circuits such as adders, multipliers, etc.

MVL circuits are confined to a small niche \cite{b2} To the best of my knowledge, there are to-day only two significant applications of MVL circuits:
\begin{itemize}
\item  Reducing the number of interconnects with multiple levels  is used in amplitude modulation: for instance, PAM-4 coding \cite{AN835}, that uses 4 levels to code 2 bits is adopted for high-speed data transmission (IEEE802.3bs). PAM-8 and PAM-16 have also been defined
\item  4-valued (MLC) flash memories store two bits per cell. 8-valued (TLC) memories store 3 bits per cell. However, these M-valued circuits (M=$2^n$) are used for higher density, not for higher speeds. 
\end{itemize}
Trying to design MVL combinational circuits to compete with binary ones looks like a dead-end.

\end{document}